\def\papertitle{Audio Transport: A Generalized Portamento Via Optimal Transport}
\def\paperauthorA{Trevor Henderson}
\def\paperauthorB{Justin Solomon}

\documentclass[twoside,a4paper]{article}
\usepackage{dafx_19}
\usepackage{amsmath,amssymb,amsfonts,amsthm}
\usepackage{euscript}
\usepackage[latin1]{inputenc}
\usepackage[T1]{fontenc}
\usepackage{ifpdf}

\usepackage[english]{babel}
\usepackage{caption}
\usepackage{subfig} 
\usepackage{color}
\usepackage{siunitx}

\ninept

\usepackage{times}

\newif\ifpdf
\ifx\pdfoutput\relax
\else
   \ifcase\pdfoutput
      \pdffalse
   \else
      \pdftrue
\fi

\ifpdf 
  \usepackage[pdftex,
    pdftitle={\papertitle},
    pdfauthor={\paperauthorA, \paperauthorB},
    colorlinks=false, 
    bookmarksnumbered, 
    pdfstartview=XYZ, 
    hidelinks
  ]{hyperref}
  \usepackage[pdftex]{graphicx}
\else 
  \usepackage[dvips]{epsfig,graphicx}
  \usepackage[dvips,
    colorlinks=false, 
    bookmarksnumbered, 
    pdfstartview=XYZ 
  ]{hyperref}
\fi
  \usepackage[hypcap=true]{caption}
\usepackage{cleveref}
\crefname{section}{\S}{\S\S}
\Crefname{section}{\S}{\S\S}
\usepackage{algorithm}
\usepackage[noend]{algpseudocode}
\title{\papertitle}

\twoaffiliations{%
\paperauthorA \,}
{\href{https://lids.mit.edu/}{Laboratory for Information and Decision Systems} \\ Massachusetts Institute of Technology \\ Cambridge, MA USA\\ {\tt \href{mailto:tfh@mit.edu}{tfh@mit.edu}}}
{\paperauthorB \,}
{\href{https://csail.mit.edu/}{Computer Science and Artificial Intelligence Laboratory} \\ Massachusetts Institute of Technology \\ Cambridge, MA USA \\ {\tt \href{mailto:jsolomon@mit.edu}{jsolomon@mit.edu}}}

\begin{document}
\ifpdf 
  \DeclareGraphicsExtensions{.png,.jpg,.pdf}
\else  
  \DeclareGraphicsExtensions{.eps}
\fi

\maketitle

\begin{abstract}
  This paper proposes a new method to interpolate between two audio signals.
  As an interpolation parameter is changed, the pitches in one signal slide to the pitches in the other, producing a portamento, or musical glide.
  The assignment of pitches in one sound to pitches in the other is accomplished by solving a 1-dimensional optimal transport problem.
  In addition, we introduce several techniques that preserve the audio fidelity over this highly nonlinear transformation.

  A portamento is a natural way for a musician to transition between notes, but traditionally it has only been possible for instruments with a continuously variable pitch like the human voice or the violin.
  Audio transport extends the portamento to \emph{any} instrument, even polyphonic ones.
  Moreover, the effect can be used to transition between different instruments, groups of instruments, or any other pair of audio signals.
%
  The audio transport effect operates in real-time; we provide an open-source implementation.
  In experiments with sinusoidal inputs, the interpolating effect is indistinguishable from ideal sine sweeps.
  More generally, the effect
  produces clear, musical results for a wide variety of inputs.
\end{abstract}

\section{Introduction}
\label{sec:intro}

A portamento, or musical glide, has been a significant expressive device in music for at least the past 200 years~\cite{leech, phonograph}.
Short portamenti can connect notes to make a passage sound more fluid, while long portamenti can draw out a transition with anticipation before finally arriving at the destination.
The author in \cite{leech} claims that ``portamento draws on innate emotional responses to human sound, as well as on our earliest memories of secure, loving communication, in order to bring to performances a sense of comfort, sincerity, and deep emotion.'' 
Regardless of whether this text describes a universal experience, portamenti have a decidedly unique sound and musical significance.

Due to the nature of the sound, the only instruments that can produce portamenti are instruments that, like the human voice, can vary their pitch continuously. 
Certain electronic systems described in~\S\ref{ssec:previous_work} are capable of producing the effect, but they are limited to particular situations (e.g.\ monophonic glide, offline processing).
In this work, we present an audio effect titled, ``audio transport,'' which interpolates between \emph{any} two audio streams in a way that sounds like a portamento, automatically and in real-time.

The audio transport effect relies on solving a 1-dimensional optimal transport problem. 
The solution to this problem determines how the pitches in one signal will move to pitches in the other. 
We find that the effect works best on pairs of sounds that do not have sharp attacks or strong tremolo and have comparable brightness.




The paper is organized as follows.
\S\ref{sec:ot_overview} gives a brief introduction to the optimal transport problem and its relevance to the rest of the paper.
\S\ref{sec:audio_transport} presents the audio transport effect, including a number of techniques necessary to produce artifact-free audio.
\S\ref{sec:experiment} details our implementation of the audio transport effect and provides perceptual results.
Finally,~\S\ref{sec:conclusion} concludes with discussion of potential applications and future work.

\subsection{Previous Work}
\label{ssec:previous_work}

Portamenti have existed in electronic music since its inception.
One of the earliest electronic instruments, the theremin, is famed for the sweeping sounds it can produce from its continuous pitch control.
Today, a pitch wheel can be found on almost all synthesizers as a way to bend a note's pitch.

In addition to manually-controlled portamenti, many synthesizers have a ``glide'' parameter which automatically introduces portamenti between sequential notes.
Typically this effect is monophonic, but some synthesizers support polyphonic glide using rule based systems~\cite{polyphonic_glide}.

As for sample-based instruments, the pitch of a sample can be changed by varying its playback speed.
Alternatively, phase vocoders allow for a sample's pitch to be changed independently of its speed~\cite{phase_vocoder, vocoder_improved}.
Both of these methods, like a pitch wheel, can produce a polyphonic portamento but they necessarily move all the pitches in the same direction at the same rate.
As such, these techniques can not be used to slide between chords with different harmonies or instruments with different timbre.

Techniques involving phase vocoders \cite{vocoder_effects}, modulation vocoders \cite{modulation_vocoder}, and popular but unpublished commercial products like Melodyne~\cite{melodyne} allow for artists to vary pitches within a sample independently, which could conceivably be used to create polyphonic portamenti.
This type of pitch manipulation, however, is not suited for real-time use because without manual input, the pitches have no destination.

While not related to portamenti, optimal transport has been applied to audio problems before.
The authors in \cite{ot_transcription} describe how optimal transport can be used to perform spectral unmixing with application to musical transcription.
The authors in~\cite{ot_pitch} apply optimal transport to the problem of fundamental pitch estimation.
Both of these papers focus on analysis rather than synthesis.

\subsection{Contributions}
\label{ssec:contributions}

We present audio transport, an audio effect that produces a portamento between arbitrary audio sources.
The effect works by interpolating between the spectra of the two input signals according to an optimal transport map.
To our knowledge, this is the first work to apply optimal transport to audio generation
and also the first that can achieve this type of portamento effect automatically and in real-time.
In addition to the novel application of optimal transport,
we present a 
technique based on time-frequency reassignment~\cite{reassignment} that divides the audio spectrum prior to transport and we extend the phase accumulation technique from~\cite{vocoder_improved} to prevent phasing between windows.



\section{Optimal Transport Overview}
\label{sec:ot_overview}

The optimal transport problem
asks how to move probability mass from one configuration to another
in a way that minimizes the amount of work (mass times distance) performed on each
infinitesimal piece of mass.
More formally~\cite{villani}, the problem seeks an optimal plan $\pi^*(x,y)$ that describes how much mass should be transferred from position $x$ to $y$ satisfying:
\begin{equation}
  \label{eqn:ot_cont}
  \pi^* =
  \underset{\pi}{\operatorname{arg\,min}}
  \iint_{\mathbb{R}^2} \|x - y\|^p\,d\pi(x, y),
\end{equation}
subject to nonnegativity as well as conservation of mass for source and target distributions $\rho_v$ and $\rho_w$:
\begin{equation}
  \int_\mathbb{R} \pi(x, y)\, dy = \rho_v(x)
  \quad\text{and}\quad
  \int_\mathbb{R} \pi(x, y)\, dx = \rho_w(y).
\end{equation}

The $p$-th root of the optimal value provides an intuitive way to measure the similarity between two distributions known as the $p$-Wasserstein distance. In the rest of this paper, we will use $p = 2$. The corresponding ``least squares'' Wasserstein distance satisfies all metric axioms among other attractive properties~\cite{cow_duck,villani}.




We use the optimal plan to perform \emph{displacement interpolation} between two distributions~\cite{displacement}.
This interpolation animates the mass assignment computed in~\Cref{eqn:ot_cont} by sliding each particle of mass between its two assignments.
In computer graphics, this interpolation technique can be used to naturally transition between histograms, images, or meshes~\cite{displacement_graphics, cow_duck, Lavenant:2018:DOT:3272127.3275064,levy2018notions,computational}.

\begin{figure}[t]
  \subfloat[A linear interpolation or ``fade'']{%
    \centerline{\includegraphics[width=\columnwidth]{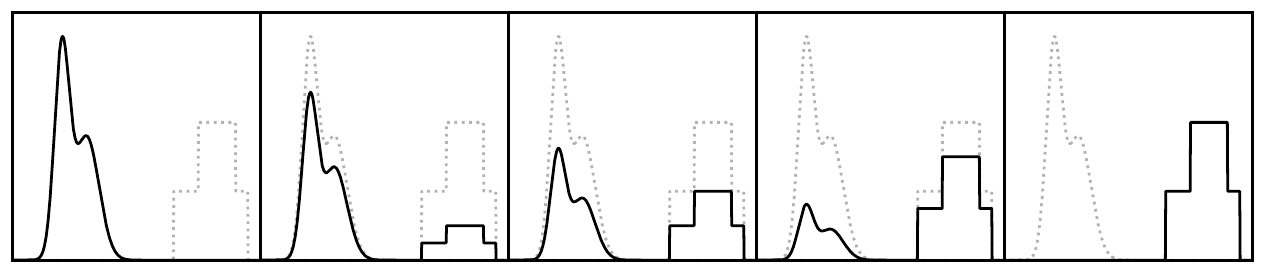}}
  }\\
  \subfloat[Displacement interpolation via optimal transport or a ``portamento'']{%
    \centerline{\includegraphics[width=\columnwidth]{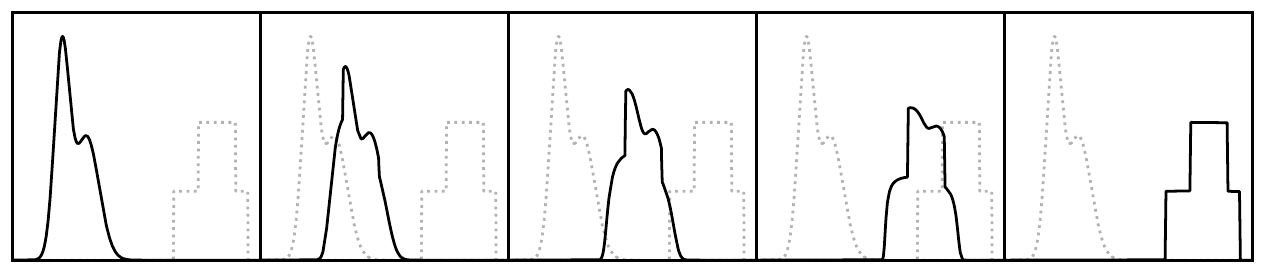}}
  }
  \caption{The distribution on the left is transformed into the distribution on the right with two different interpolation methods.}
  \label{fig:linear_vs_ot}
\end{figure}

Consider~\Cref{fig:linear_vs_ot}, 
which demonstrates two different ways to interpolate between distributions.
On top, the distributions are interpolated linearly.
If we imagine the distributions as audio spectra, then this transformation is simply fading one set of pitches out and another set in.
On the bottom, the same distributions are transformed using displacement interpolation.
The mass physically slides from one location to another.
If these were audio spectra, this sliding would sound like a portamento.

It should be noted that solving the optimal transport problem is known to be computationally challenging for any dimension $d > 1$. 
Fortunately, solving the problem on the real line can be done in linear time~\cite{computational}.

\section{Audio Transport}
\label{sec:audio_transport}

The audio transport effect works by performing displacement interpolation on input audio spectra, so that pitches in one signal slide to pitches in the other as an interpolation parameter is changed.
To modify the spectra over time, the audio transport algorithm follows the phase-vocoder paradigm~\cite{phase_vocoder, vocoder_improved, vocoder_effects}.
In detail, a sliding short-time Fourier transform (STFT) is applied to both input audio streams, producing complex spectra.
These spectra are interpolated according to the optimal transport map and fed through an  
inverse STFT to form the output audio stream.

\S\ref{ssec:transport} describes an efficient way to interpolate between spectra using optimal transport.
Alone, this method produces two artifacts which, borrowing from phase-vocoder literature, are known as \emph{vertical incoherence} and \emph{horizontal incoherence}~\cite{vocoder_improved}.
Solutions to these two phenomena are described in~\S\ref{ssec:vertical} and \S\ref{ssec:horizontal}, respectively.



\subsection{Optimal Transport Between Spectra}
\label{ssec:transport}

Consider discrete spectra represented by complex vectors $X$, $Y$ and corresponding frequency vectors $\omega^X$, $\omega^Y$.
Analogously to the continuous optimal transport plan given in~\Cref{eqn:ot_cont}, we can write the optimal transport plan between these discrete spectra
as the plan $\pi^*\in \mathbb{R}^{|X|\times|Y|}$ minimizing:
\begin{equation}
  \label{eqn:ot_spectral}
  \pi^* =
  \underset{\pi\geq0}{\operatorname{arg\,min}}
  \sum_{i,j} \left|\omega^X_i - \omega^Y_j\right|^2\pi_{ij}
\end{equation}
subject to the conservation of mass constraint
\begin{equation}
  \label{eqn:ot_spectral_constraint}
  \sum_j \pi_{ij} = |X_i|
  \quad\text{and}\quad
  \sum_i \pi_{ij} = |Y_j|.
\end{equation}
This problem assumes that $\sum_i |X_i| = \sum_j |Y_j|$. To treat spectra with different total magnitudes, the plan can be computed on normalized spectra; then, scaling is interpolated linearly over the interpolation.

Once an optimal plan is computed, the spectra can be interpolated with parameter $k\in[0,1]$ by placing each mass $\pi^*_{ij}$ at the displaced frequency:
\begin{equation}
  \label{eqn:interpolated_freq}
  (1 - k)\omega_i^X + k\omega_j^Y
\end{equation}
If multiple masses are placed at the same frequency, they are added together.
The phase attributed to the mass is considered in~\S\ref{ssec:horizontal}.

In one dimension, the optimal transport plan is monotone or, in other words, no mass crosses over any other mass~\cite{applied}.
This allows for~\Cref{eqn:ot_spectral,eqn:ot_spectral_constraint} to be solved using the greedy strategy presented in~\Cref{alg:ot}. 

The algorithm begins with the initial bins of the two spectra.
Since no mass can cross over any other,
all of the mass in the smaller bin must be assigned to the larger.
With this assignment done, one can imagine virtually removing the smaller bin and shrinking the mass of the larger by the mass assignment.
The algorithm then continues inductively on the smaller problem. 
%
At every iteration, all of the mass in one bin becomes completely assigned.
Therefore, the complexity of the algorithm is $O(|X| + |Y|)$.
This runtime is efficient relative to the super-linear runtime of the fast Fourier transform.

\begin{algorithm}[t]
  \caption{Computing The Optimal Transport Matrix, $\pi^*$}\label{alg:ot}
\begin{algorithmic}
  \State{$\pi^*_{i,j} \leftarrow 0$}
  \State{$\rho_X, \rho_Y \leftarrow |X_0|, |Y_0|$}
  \Comment{$\rho$ is the mass left in a bin}
  \\
  \Loop
    \If{$\rho_X < \rho_Y$}
      \State{$\pi^*_{ij} \leftarrow \rho_X$}
      \Comment{Assign as much mass as possible}
      \\
      \State{$i \leftarrow i+1$}
      \Comment{Refill the emptied bin}
      \If{$i \geq |X|$} break\EndIf
      \State{$\rho_X \leftarrow |X_i|$}
      \\
      \State{$\rho_Y \leftarrow \rho_Y - \rho_X$}
      \Comment{Decrease the capacity of the other}
      \\
    \Else
      \State{Symmetric to the case above}
    \EndIf
  \EndLoop
  \\\\
  \Return{$\pi^*$}
\end{algorithmic}
\end{algorithm}

\subsection{Resolving Vertical Incoherence: Slicing the Spectrogram}
\label{ssec:vertical}

\begin{figure}[t]
  \centerline{\includegraphics[width=\columnwidth]{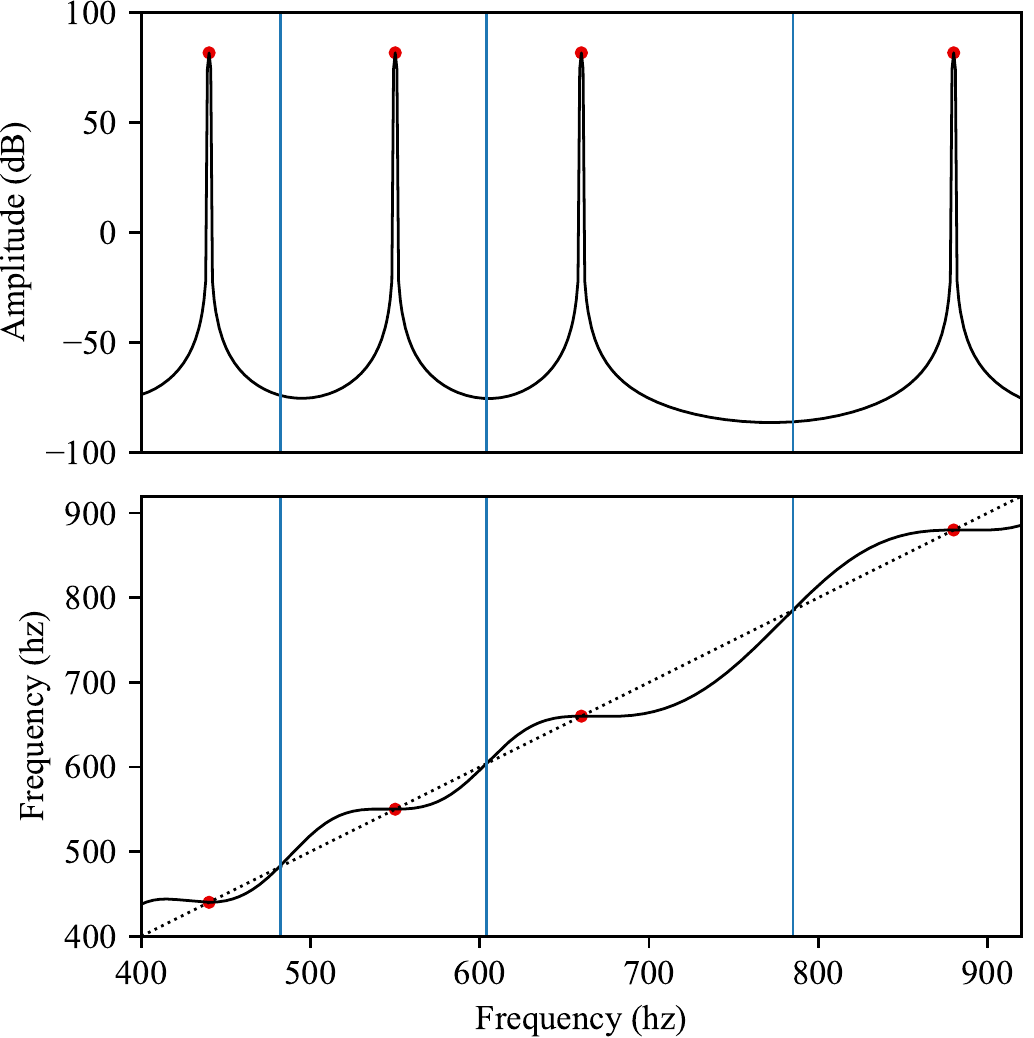}}
  \caption{Dividing the spectrum of a sinusoidal A major chord consisting of the notes A$_4$, C$\sharp_5$, E$_5$ and A$_5$. The spectrum is displayed on top. On the bottom, the reassigned frequency $\hat{\omega}$ (solid line) is plotted against the frequency $\omega$ (dashed line). The intersections of these lines indicate the boundaries between groups (vertical lines) and their pitch centers (dots).}
  \label{fig:groups}
\end{figure}

One unfortunate effect of using an STFT is the necessary trade-off between time and frequency resolution.
As the time resolution increases, the frequency domain becomes ``smeared.''

The relation between a peak frequency and its smeared components is known to be important for perceptual quality.
Treating these independently leads to phasing artifacts within a window known as vertical incoherence~\cite{vocoder_improved}.
One method to solve this problem in phase vocoder literature
is to ``lock'' regions surrounding a peak frequency so that the relative phase between bins within these regions remains unchanged~\cite{phase_lock, vocoder_improved}.

If~\Cref{alg:ot} were applied directly to audio spectra, it would introduce vertical incoherence by translating smeared components independently.
So, applying the locking strategy, we will treat smeared regions as single units with collective magnitude in the transportation map.

It now remains to determine how exactly to choose the boundaries between smeared spectral regions.
A common strategy is to use a heuristic to find local peaks and then assign the boundaries to be the midpoints of the peaks.
Since displacement interpolation makes extreme changes to the spectra, however,
this somewhat na\"ive plan~\cite{vocoder_improved, vocoder_effects} is not sufficiently robust to produce a clean signal.
We propose a more principled segmentation method based on frequency reassignment.

Frequency reassignment uses information in a signal's phase to enhance its frequency resolution. 
Each spectral component with frequency $\omega_i$ is mapped to the reassigned frequency $\hat{\omega}_i$ that better reflects the true energy distribution~\cite{reassignment}. 
Sinusoids that have been smeared across multiple bins become mapped to the same central frequency, which produces the plateaus shown in~\Cref{fig:groups}.

With this view, an intuitive way to define sinusoidal regions is by the zero crossings of $\hat{\omega}_i - \omega_i$.
Falling crossings indicate the center bin of a region while rising crossings indicate the boundaries.
These can be computed at the cost of an additional STFT with the following formula~\cite{reassignment}:

\begin{align}
  \hat{\omega}_i - \omega_i = \Im\left\{\frac{X^{\mathcal{T}h}_i \cdot X^*_i}{|X_i^2|}\right\}.
\end{align}
$X^{\mathcal{T}h}$ is the STFT computed using a time-weighted analysis window.

\subsection{Resolving Horizontal Incoherence: Phase Accumulation}
\label{ssec:horizontal}

Finally, we reintroduce phase to the spectra.
In doing so, we will be concerned with the phase relations between consecutive windows rather than the phase relations within a window.
Inter-window phase relations
carry information about short-time events like transients
and hence ignoring these relations can create a blurry sound in some cases as discussed in \Cref{sec:natural}.

When a particular spectral region is transposed, its phase rotates at a different rate.
Thus, applying the phases of consecutive windows in the original signal to the corresponding windows of the transposed signal causes interference known as horizontal incoherence~\cite{vocoder_improved}.

In phase vocoders, this is resolved by integrating the reassigned frequency over the window difference.
In other words, the phase $\varphi_{i}^t$ in bin $i$ and window $t$ can be estimated from the phase $\varphi_i^{t-1}$ in window $t-1$ as follows:
\begin{align}
  \label{eqn:phase_accumulation}
  \varphi_{i}^t = \varphi_i^{t-1} + \hat{\omega}_i^{t-1}\cdot \Delta,
\end{align}
where $\Delta$ is the delay between the windows in seconds.
This update is applied to \emph{center} bin of a region as described in~\S\ref{ssec:vertical}.
The other phases in each region are modified accordingly to maintain the same relative phase with respect to the center bin.

Some small modifications must be made to apply \Cref{eqn:phase_accumulation} to the audio transport effect.
First of all, it is possible that many spectral regions overlap on a particular bin, which makes the terms $\hat{\omega}_i^{t-1}$ and $\varphi_i^{t-1}$ ambiguous. To resolve this we simply choose the frequency and phase of the loudest overlapping region. 
Additionally, since the audio transport effect consists of rapidly-moving pitches, we can minimize phasing by averaging the current reassigned frequency and the previous reassigned frequency:
\begin{align}
  \varphi_{i}^t = \varphi_i^{t-1} + \frac{\hat{\omega}_i^t + \hat{\omega}_i^{t-1}}{2}\cdot \Delta.
\end{align}



\section{Results}
\label{sec:experiment}

We implemented the audio transport effect described in \S\ref{sec:audio_transport} for real-time audio interpolation.
We tested our implementation on synthetic sounds described in \S\ref{sec:sines} as well as on a variety of complex and natural sounds described in \S\ref{sec:natural}. 

All of our results are performed on \SI{44.1}{\kilo\hertz} audio with a window size of \SI{0.05}{\second} or 2206 samples.
We use a Hann analysis window with $50\%$ overlap and no synthesis window.
Additionally, the windows are padded with zeros to increase the frequency resolution of the FFT to $\approx\SI{5}{\hertz}$.

Our implementation is open source and available at
\url{https://github.com/sportdeath/audio_transport}.

\subsection{Interpolating Sinusoids}
\label{sec:sines}

\begin{figure}[t]
  \centerline{\includegraphics[width=\columnwidth]{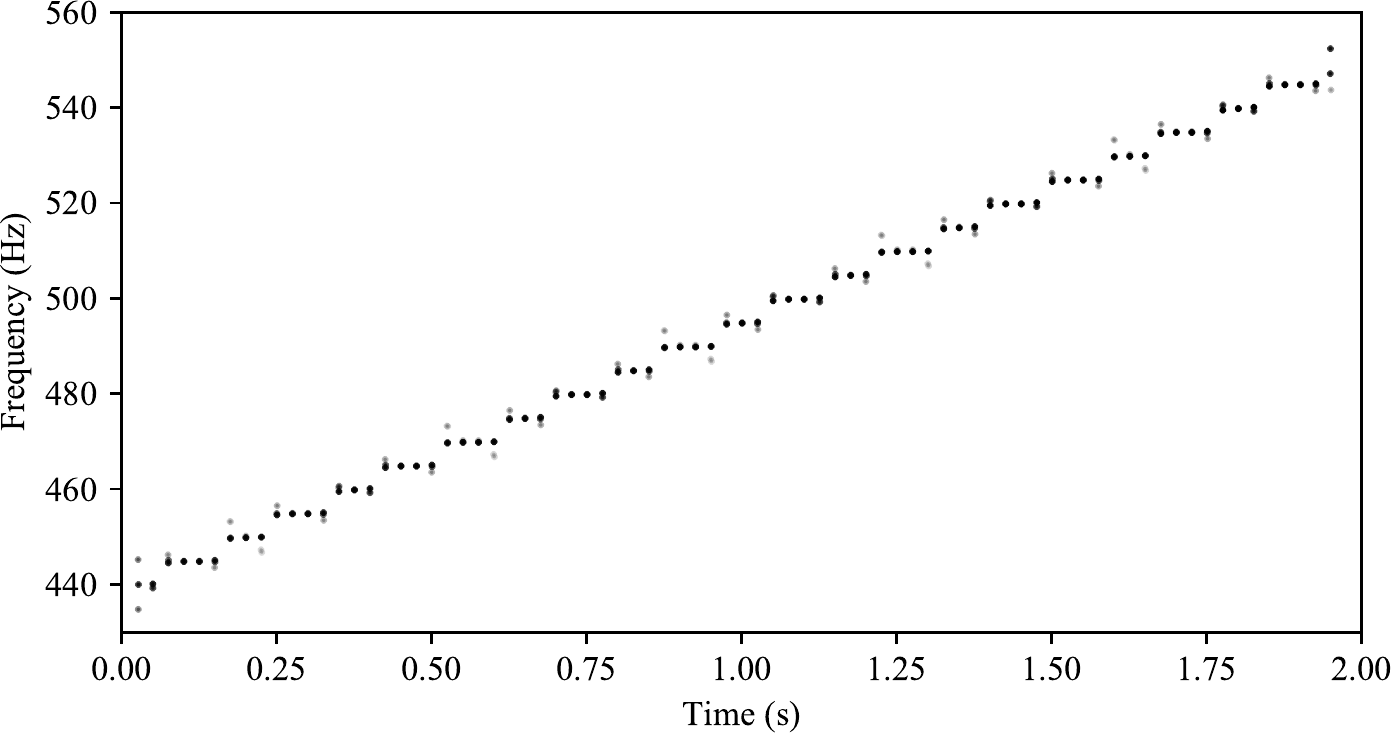}}
\caption{A (reassigned~\cite{reassignment}) spectrogram of the audio transport effect being used to interpolate between sinusoids at A$_4$ and C$\sharp_5$. The ``stair-stepping'' effect is due to the frequency resolution of $\approx\SI{5}{\hertz}$, but this can be reduced arbitrarily by padding each window with additional zeros.}
  \label{fig:sine_spec}
\end{figure}

We used the audio transport effect to interpolate between single sinusoids.
Intuitively, this should sound exactly like a sine sweep between the input pitches.
We performed listening experiments for interpolations at a variety of speeds and with inputs spanning the entire perceptive range.
The spectrogram of one such interpolation is shown in~\Cref{fig:sine_spec}.
Almost all of the interpolations were indistinguishable from real sine sweeps.
In extreme cases where the interpolations were faster than $\SI{2000}{\hertz\per\second}$ we perceived some phase distortion, but these situations would be rare in normal use.

The audio does exhibit ``stair-stepping'' between frequencies due to the frequency resolution of the FFT and the time resolution of the windows as demonstrated in~\Cref{fig:sine_spec}.
Due to the small time-frequency resolution of the steps, however, we were unable to perceive them in the listening experiments.

\subsection{Interpolating Natural Sounds}
\label{sec:natural}

\begin{figure}[t]
  \centerline{\includegraphics[width=\columnwidth]{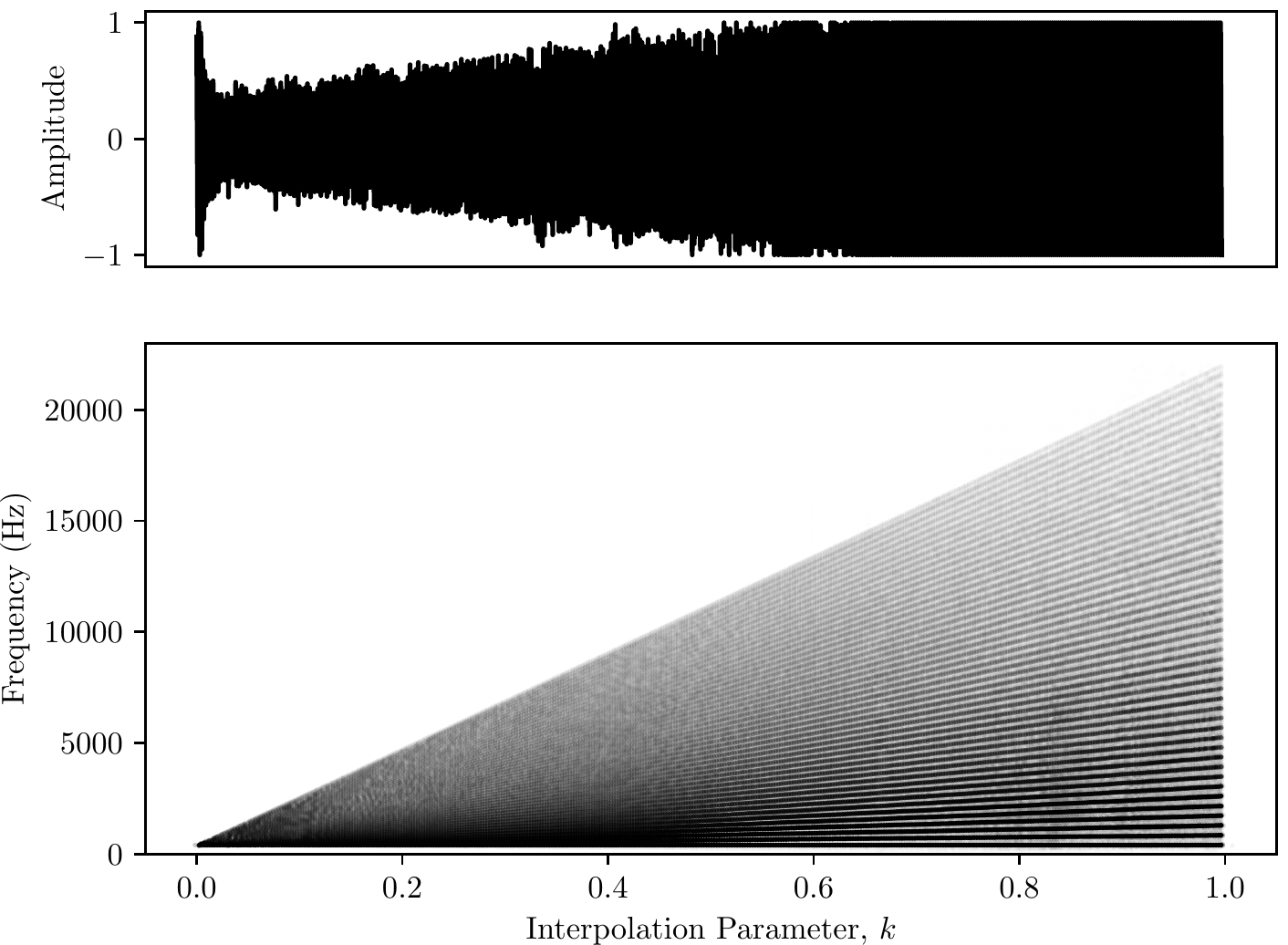}}
  \caption{Interpolating between a sine wave ($k = 0$) and a saw wave ($k =1$) leads to a drop in volume for values of $k$ close to but not equal to zero as shown in the audio signal on top.
    The spectrogram shows how the sine's single peak splits into many peaks, which interfere when they are close together.
  }
  \label{fig:sine_saw}
\end{figure}

We tested the audio transport effect on a variety of sounds, some of which are available at 
\url{https://soundcloud.com/audio_transport}.
The effect intuitively sounds like a portamento, even when applied to unnatural cases like a piano note gliding into a human voice.
It is applicable to monophonic and polyphonic sounds. We even had success using it to transition between entire songs.
The audio typically sounds artifact-free for many classes of audio signals, with a few exceptions noted below.

The audio transport effect does not guarantee temporal consistency between transport maps. So interpolating between sounds with dynamic spectra, like a pair of wavering orchestral chords, can produce a fluctuating pitch.

Another artifact that can occur is a sudden drop in volume when the interpolation parameter is close to but not equal to either zero or one as shown in \Cref{fig:sine_saw}.
This happens when one single frequency is mapped to a large range of frequencies.
As the single frequency separates, its components interfere with each other, reducing the volume.
This artifact is most prevalent when interpolating between sounds with vastly different spectral complexity as is the case with pairs of bright and dark sounds.

It is also worth reminding the reader that this method is intended for static sounds and will blur transients, as mentioned in \S\ref{ssec:horizontal}.
This artifact can be subtle, but when we directly compared the output of the audio transport effect with an interpolation factor of $0$ to the corresponding input we consistently picked out the original audio when it had sharp transients like hi-hats.
We suspect that this could be fixed using phase reinitialization techniques~\cite{vocoder_improved}, but this exploration is left to future work.

\section{Conclusion}
\label{sec:conclusion}

In this paper, we introduced the \emph{audio transport} effect which can create a portamento-like transition between any two audio signals. 
The effect produces a novel but intuitive sound and it is controlled by a single interpolation parameter.
As a result, it is accessible for musicians to incorporate into both live performances and studio recordings.

In our live experiments, we controlled the interpolation parameter using a MIDI pitch fader, but really it can come from any source. 
For example, an instrument could be constructed where the velocity of a note controls the interpolation parameter.
As much as we have described the effect as a portamento, the input pitches do not need to have a different fundamental. The effect also produces interesting interpolations between signals with the same fundamental pitch but different timbre.

Our work on audio transport suggests several
other use cases beyond those explored in our experiments. 
For example, consider a single audio source that is fed as one input of the audio transport effect, and the output of the effect is fed back into the other input.
By keeping the interpolation parameter constant, the pitches in the output should lag behind the input pitches similar to synthesizer ``glide.'' This setup leads to several questions:  
What would happen if other effects were added to the feedback chain? Is it interesting to use multiple audio transport effects at the same time?  The latter may be supported by the notion of \emph{barycenters} in optimal transport~\cite{agueh2011barycenters}.



The audio transport effect as described still produces artifacts for certain classes of sounds.
Future work to resolve these could investigate ways to sharpen transients, make transport maps temporally consistent, and reduce the effects of energy cancellation.
For a wide variety of inputs, however, our effect sounds smooth, musical and inspiring.




\paragraph*{Acknowledgments.} The authors acknowledge the generous support of Army Research Office grant W911NF-12-R-0011, of Air Force Office of Scientific Research award FA9550-19-1-0319, of National Science Foundation grant IIS-1838071, from an Amazon Research Award, and from the MIT-IBM Watson AI Laboratory. Any opinions, findings, and conclusions or recommendations expressed in this material are those of the authors and do not necessarily reflect the views of these organizations.

\nocite{*}
\bibliographystyle{IEEEbib}
\bibliography{bib} 

\begin{thebibliography}{10}

\bibitem{leech}
Daniel Leech-Wilkinson,
\newblock ``Portamento and musical meaning,''
\newblock {\em Journal of Musicological Research}, vol. 25, no. 3-4, pp.
  233--261, 2006.

\bibitem{phonograph}
Mark Katz,
\newblock ``Portamento and the phonograph effect,''
\newblock {\em Journal of Musicological Research}, vol. 25, no. 3-4, pp.
  211--232, 2006.

\bibitem{polyphonic_glide}
Ralph Deutsch and Leslie~J Deutsch,
\newblock ``Constant speed polyphonic portamento system,'' Oct.~19 1982,
\newblock US Patent 4,354,414.

\bibitem{phase_vocoder}
James~L. Flanagan and R.M. Golden,
\newblock ``Phase vocoder,''
\newblock {\em Bell System Technical Journal}, vol. 45, no. 9, pp. 1493--1509,
  1966.

\bibitem{vocoder_improved}
Jean Laroche and Mark Dolson,
\newblock ``Improved phase vocoder time-scale modification of audio,''
\newblock {\em IEEE Transactions on Speech and Audio Processing}, vol. 7, no.
  3, pp. 323--332, 1999.

\bibitem{vocoder_effects}
Jean Laroche and Mark Dolson,
\newblock ``New phase-vocoder techniques for pitch-shifting, harmonizing and
  other exotic effects,''
\newblock in {\em Workshop on Applications of Signal Processing to Audio and
  Acoustics}. IEEE, 1999, pp. 91--94.

\bibitem{modulation_vocoder}
Sascha Disch and Bernd Edler,
\newblock ``An amplitude-and frequency modulation vocoder for audio signal
  processing,''
\newblock in {\em Proc. of the Int. Conf. on Digital Audio Effects (DAFx)},
  2008.

\bibitem{melodyne}
Peter Neubaecker,
\newblock ``Sound-object oriented analysis and note-object oriented processing
  of polyphonic sound recordings,'' 2009,
\newblock US8022286B2.

\bibitem{ot_transcription}
R{\'e}mi Flamary, C{\'e}dric F{\'e}votte, Nicolas Courty, and Valentin Emiya,
\newblock ``Optimal spectral transportation with application to music
  transcription,''
\newblock in {\em Advances in Neural Information Processing Systems}, 2016, pp.
  703--711.

\bibitem{ot_pitch}
Filip Elvander, Stefan~Ingi Adalbj{\"o}rnsson, Johan Karlsson, and Andreas
  Jakobsson,
\newblock ``Using optimal transport for estimating inharmonic pitch signals,''
\newblock in {\em International Conference on Acoustics, Speech and Signal
  Processing (ICASSP)}. IEEE, 2017, pp. 331--335.

\bibitem{reassignment}
Kelly~R Fitz and Sean~A Fulop,
\newblock ``A unified theory of time-frequency reassignment,''
\newblock {\em arXiv:0903.3080}, 2009.

\bibitem{villani}
C{\'e}dric Villani,
\newblock {\em Optimal Transport: Old and New}, vol. 338,
\newblock Springer Science \& Business Media, 2008.

\bibitem{cow_duck}
Justin Solomon, Fernando De~Goes, Gabriel Peyr{\'e}, Marco Cuturi, Adrian
  Butscher, Andy Nguyen, Tao Du, and Leonidas Guibas,
\newblock ``Convolutional {W}asserstein distances: Efficient optimal
  transportation on geometric domains,''
\newblock {\em ACM Transactions on Graphics (TOG)}, vol. 34, no. 4, pp.
  66:1--66:11, 2015.

\bibitem{displacement}
Robert~J McCann,
\newblock ``A convexity principle for interacting gases,''
\newblock {\em Advances in Mathematics}, vol. 128, no. 1, pp. 153--179, 1997.

\bibitem{displacement_graphics}
Nicolas Bonneel, Michiel Van De~Panne, Sylvain Paris, and Wolfgang Heidrich,
\newblock ``Displacement interpolation using {L}agrangian mass transport,''
\newblock in {\em ACM Transactions on Graphics (TOG)}. ACM, 2011, vol.~30, pp.
  158:1--158:12.

\bibitem{Lavenant:2018:DOT:3272127.3275064}
Hugo Lavenant, Sebastian Claici, Edward Chien, and Justin Solomon,
\newblock ``Dynamical optimal transport on discrete surfaces,''
\newblock {\em ACM Trans. Graph.}, vol. 37, no. 6, pp. 250:1--250:16, Dec.
  2018.

\bibitem{levy2018notions}
Bruno L{\'e}vy and Erica~L Schwindt,
\newblock ``Notions of optimal transport theory and how to implement them on a
  computer,''
\newblock {\em Computers \& Graphics}, vol. 72, pp. 135--148, 2018.

\bibitem{computational}
Gabriel Peyr{\'e}, Marco Cuturi, et~al.,
\newblock ``Computational optimal transport,''
\newblock {\em Foundations and Trends in Machine Learning}, vol. 11, no. 5-6,
  pp. 355--607, 2019.

\bibitem{applied}
Filippo Santambrogio,
\newblock {\em Optimal Transport for Applied Mathematicians}, vol.~55,
\newblock Springer, 2015.

\bibitem{phase_lock}
Miller Puckette,
\newblock ``Phase-locked vocoder,''
\newblock in {\em Workshop on Applications of Signal Processing to Audio and
  Accoustics}. IEEE, 1995, pp. 222--225.

\bibitem{agueh2011barycenters}
Martial Agueh and Guillaume Carlier,
\newblock ``Barycenters in the {W}asserstein space,''
\newblock {\em SIAM Journal on Mathematical Analysis}, vol. 43, no. 2, pp.
  904--924, 2011.

\end{thebibliography}

\end{document}